\documentclass{acm_sen_article}

\usepackage{todonotes}
\usepackage{lipsum}
\usepackage{flushend}
\RequirePackage{doi}
\usepackage{hyperref}
\usepackage{paralist}

\usepackage[numbers]{natbib} 

\begin{document}

\title{Greening AI-enabled Systems with Software Engineering: A Research Agenda for Environmentally Sustainable AI Practices}

\numberofauthors{29} 
\author{
       Luís Cruz\\
       \affaddr{Delft University of Technology}\\
       \affaddr{Delft, The Netherlands}\\
       \email{L.Cruz@tudelft.nl}
\and 
João Paulo Fernandes\\
       \affaddr{New York University Abu Dhabi}\\
       \affaddr{United Arab Emirates}\\
       \email{jpf9731@nyu.edu}
\and
Maja H.\ Kirkeby\thanks{Corresponding author.}\\
       \affaddr{Roskilde University}\\
       \affaddr{City, Country}\\
       \email{kirkebym@acm.org}
\and
Silverio Martínez-Fernández\\
       \affaddr{Universitat Politècnica de Catalunya}\\
       \affaddr{Barcelona, Spain}\\
       \email{silverio.martinez@upc.edu}
\and 
June Sallou\thanks{The first five authors (alphabetic order) served as workshop organizers. The remaining authors (alphabetic order) participated in the workshop and actively contributed to the writing and review of the article.}\\
       \affaddr{Wageningen University \& Research}\\
       \affaddr{The Netherlands}\\
       \email{june.sallou@wur.nl}
\and
%
%
%
%
%
Hina Anwar\\
    \affaddr{University of Tartu}\\
    \affaddr{Tartu, Estonia}\\
    \email{hina.anwar@ut.ee}
\and
Enrique Barba Roque\\
    \affaddr{Delft University of Technology}\\
    \affaddr{Delft, The Netherlands}\\
    \email{e.barbaroque@tudelft.nl}
\and
Justus Bogner\\
    \affaddr{Vrije Universiteit Amsterdam}\\
    \affaddr{Amsterdam, The Netherlands}\\
    \email{j.bogner@vu.nl}
\and
Joel Castaño\\
    \affaddr{Universitat Politècnica de Catalunya}\\
    \affaddr{Barcelona, Spain}\\
    \email{joel.castano@upc.edu}
\and
Fernando Castor\\
    \affaddr{University of Twente}\\
    \affaddr{Enschede, The Netherlands}\\
    \email{f.castor@utwente.nl}
\and
Aadil Chasmawala\\
    \affaddr{New York University Abu Dhabi}\\
    \affaddr{United Arab Emirates}\\
    \email{aac10066@nyu.edu}
\and
Simão Cunha\\
    \affaddr{University of Minho}\\
    \affaddr{Braga, Portugal}\\
    \email{a93262@alunos.uminho.pt}
\and
Daniel Feitosa\\
    \affaddr{University of Groningen}\\
    \affaddr{Groningen, The Netherlands}\\
    \email{d.feitosa@rug.nl}
\and
Alexandra González\\
    \affaddr{Universitat Politècnica de Catalunya}\\
    \affaddr{Barcelona, Spain}\\
    \email{alexandra.gonzalez.alvarez@upc.edu}
\and
Andreas Jedlitschka\\
    \affaddr{Fraunhofer Institute for Experimental Software Engineering}\\
    \affaddr{Kaiserslautern, Germany}\\
    \email{Andreas.Jedlitschka@iese.fraunhofer.de}
\and
Patricia Lago\\
    \affaddr{Vrije Universiteit Amsterdam}\\
    \affaddr{Amsterdam, The Netherlands}\\
    \email{p.lago@vu.nl}
\and
    Henry Muccini\\
     \affaddr{University of L'Aquila}\\
    \affaddr{L'Aquila, Italy}\\
    \email{henry.muccini@univaq.it}
\and
Ana Oprescu\\
    \affaddr{Universiteit van Amsterdam}\\
    \affaddr{Amsterdam, The Netherlands}\\
    \email{a.m.oprescu@uva.nl}
\and
Pooja Rani \\
    \affaddr{University of Zurich}\\
    \affaddr{Zurich, Switzerland}\\
    \email{rani@ifi.uzh.ch}
\and
João Saraiva \\
    \affaddr{University of Minho \& INESC TEC}\\
    \affaddr{Braga, Portugal}\\
    \email{saraiva@di.uminho.pt}
\and
Federica Sarro \\
    \affaddr{University College London}\\
    \affaddr{London, United Kingdom}\\
    \email{f.sarro@ucl.ac.uk}
\and
Raghavendra Selvan\\
    \affaddr{University of Copenhagen}\\
    \affaddr{Copenhagen, Denmark}\\
    \email{raghav@di.ku.dk}
\and
Karthik Vaidhyanathan\\
    \affaddr{SERC, IIIT Hyderabad}\\
    \affaddr{Hyderabad, India}\\
    \email{karthik.vaidhyanathan@iiit.ac.in}
\and
Roberto Verdecchia\\
    \affaddr{University of Florence}\\
    \affaddr{Florence, Italy}\\
    \email{roberto.verdecchia@unifi.it}
\and
     Ivan P. Yamshchikov\\
    \affaddr{THWS}\\
    \affaddr{Würzburg, Germany}\\
    \email{ivan.yamshchikov@thws.de}
}


\maketitle

\begin{abstract}
The environmental impact of Artificial Intelligence (AI)-enabled systems is increasing rapidly, and software engineering plays a critical role in developing sustainable solutions.
The ``Greening AI with Software Engineering'' workshop,\footnote{\href{https://www.cecam.org/workshop-details/greening-ai-with-software-engineering-1358}{https://www.cecam.org/workshop-details/greening-ai-with-software-engineering-1358}} funded by the \textit{Centre Européen de Calcul Atomique et Moléculaire} (\href{https://www.cecam.org/}{CECAM}) and the \href{https://www.lorentzcenter.nl/}{Lorentz Center}, provided an interdisciplinary forum for 29 participants, from practitioners to academics, to share knowledge, ideas, practices, and current results dedicated to advancing green software and AI research. The workshop was held February 3-7, 2025, in Lausanne, Switzerland. Through keynotes, flash talks, and collaborative discussions, participants identified and prioritized key challenges for the field. These included {\em energy assessment and standardization}, {\em benchmarking practices}, {\em sustainability-aware architectures}, {\em runtime adaptation}, {\em empirical methodologies}, and {\em education}. This report presents a research agenda emerging from the workshop, outlining open research directions and practical recommendations to guide the development of environmentally sustainable AI-enabled systems rooted in software engineering principles.


\end{abstract}

\section{Introduction}

The rapid advancement of Artificial Intelligence (AI) has brought substantial benefits across numerous domains, but it has also raised growing concerns about its environmental sustainability. As AI models increase in size, complexity, and deployment scale, so do their energy consumption and carbon footprint (e.g., \cite{Luccioni2024}). Training large-scale models can consume as much energy as powering homes for weeks or months, while inference at scale further compounds this impact. Yet methods for measuring, optimizing, and communicating these effects remain fragmented, inconsistent, and underdeveloped \cite{MenziesIEEEsw24}.

Software Engineering (SE) has a critical role to play in addressing these pressing challenges. Architectural decisions, development practices, tooling, benchmarking, and lifecycle management, all shape the environmental profile of AI-enabled systems. However, sustainable AI development is inherently interdisciplinary, requiring coordination across AI research, systems engineering, education, and policy. To understand how SE can contribute meaningfully to this space, we must examine methodological practices, educational strategies, tools, and infrastructure-level design. 
%
%
As such this article is a response to the call for contributions to  address emerging trends from an SE perspective~\cite{SEN_aim}. 

This article presents key insights from the ``Greening AI with Software Engineering'' workshop held in Lausanne, Switzerland, on February 3-7, 2025, co-financed by the Centre Européen de Calcul Atomique et Moléculaire (CECAM) and the Lorentz Center.
The workshop gathered 29 participants from academia and industry for an intensive program of keynotes, flash talks, and breakout discussions. Designed to support co-creation, the workshop format enabled participants to collaboratively identify and refine the focus areas that structure this report.\footnote{These focus areas emerged through an iterative process: The participants first contributed individual reflections using sticky notes, which were grouped into preliminary clusters. These were refined through rotating group discussions supported by anchors who maintained continuity and captured insights. Then, the participants committed to a single area and co-developed draft texts. After the workshop, the coordinators consolidated the material, removed underdeveloped topics, and identified cross-cutting themes. A draft report was shared with the participants and revised based on their feedback.} In doing so, the workshop surfaced open questions, practical tensions, and recurring patterns in how AI sustainability is approached across domains.

The five focus areas presented in this report are: \begin{inparaenum}[1)] \item energy assessment and standardization, \item evaluation and benchmarking of AI sustainability, \item software architecture and lifecycle, \item empirical methods and reproducibility, and \item education and awareness. \end{inparaenum} These areas structure the main body of the article and reflect both the breadth and depth of sustainability challenges discussed during the workshop.
Across the five focus areas, we identify cross-cutting concerns such as {\em standardization}, {\em metric design}, and {\em holistic thinking} that shape the broader research agenda. By summarizing and synthesizing these discussions, this article contributes to building a shared vocabulary of key terms, distinctions, and recurring themes, and outlines a research agenda for advancing environmentally sustainable AI with Software Engineering. A particular effort has been made to emphasize how we imagine Software Engineering can contribute across these five areas.

\section{Energy Assessment and Standardization}
Accurately assessing the energy consumption of AI-enabled systems is a foundational step toward understanding and improving their environmental impact. However, the current landscape of energy measurement and estimation is fragmented, with a lack of or inconsistent tools, granularity levels, and reporting practices. This section outlines the main challenges associated with measuring energy usage in AI-enabled systems.

\subsection{Bridging Granularity and Standardization in Energy Metrics}

Achieving high-level standards for collecting system-wide energy footprint data requires a standardized approach similar to the Intel’s Running Average Power Limit (RAPL)\footnote{\url{https://www.intel.com/content/www/us/en/developer/articles/technical/software-security-guidance/advisory-guidance/running-average-power-limit-energy-reporting.html}}, but one that applies uniformly across all computational devices.
Standardization efforts prioritize broad applicability across entire software systems rather than precision at the component level. However, organizations require a more granular perspective on energy consumption to support informed decision-making.

Establishing high-level standards would enable the collection of consistent and comparable metrics across diverse computing environments. A holistic methodology should capture energy consumption data at multiple granularities, from system-wide reporting to fine-grained scientific measurements, ensuring traceability across levels.
A standardized approach should facilitate comparing different solutions, enabling consistent evaluation of their resource demands across computing devices.
It should support precise measurements when needed, such as analyzing short-lived executions. 

These objectives align with ongoing research on energy consumption in AI and Machine Learning (ML), as seen in studies such as Green AI \cite{schwartz2020green}, Carbontracker \cite{anthony2020carbontracker}, and research on estimating the carbon footprint of large-scale AI models~\cite{Luccioni2023}.

\subsection{Methodological Consistency}

Assessing energy consumption of an ML model is inherently challenging because a model's behavior depends on multiple factors, including architecture, platform, hyperparameters, quantization level, task type, and input/output characteristics. Additionally, ML models exhibit non-deterministic execution, making reproducibility a significant challenge. Since testing every possible combination of influencing variables is infeasible, identifying key sources of variation is essential to obtain meaningful and transferable results.


Unlike program execution time, the energy consumption of ML models is difficult to explain. Without robust methodologies, assessments risk being inaccurate, inconsistent, or non-generalizable across different contexts. Ensuring accuracy, reliability, and reproducibility is crucial to making energy measurements reflect real-world behavior as closely as possible.


Addressing this challenge requires balancing abstraction and precision -- developing a methodology that is both widely applicable and detailed enough to capture ML models' energy behavior accurately. This involves distinguishing between factors that significantly impact energy consumption and those that can be abstracted away without compromising assessment validity. For instance, on the one hand, studies suggest that in code generation with large language models, output length influences energy consumption more than input length \cite{alizadeh2024language}. On the other hand, in tasks such as code completion, the large amount of context required to generate small pieces of code may result in a shift in this balance. Another promising approach is using mini-models as proxies for larger models, leveraging evidence that smaller models with the same architecture can effectively estimate the energy footprint of their larger counterparts \cite{qiu2024power}.


Due to the heterogeneity and continuous evolution of AI-enabled systems, a single, rigid methodology for energy assessment is inadequate. The diversity of models, architectures, hardware, and workloads prevents any universal approach from accurately capturing energy consumption in all cases. Moreover, as AI technologies evolve, current methodologies may become obsolete. This presents two key challenges: first, providing practitioners with reliable guidance in selecting appropriate energy assessment methods, and second, ensuring that methodologies remain adaptable as AI-enabled systems continue to change.



To address these challenges, a flexible yet structured approach is essential. Instead of enforcing a single assessment framework, practitioners need adaptable guidelines that enable them to navigate the trade-offs between different methodologies while ensuring compliance with high-level energy assessment standards. One potential solution is the creation of a {\em publicly accessible repository} of energy assessment methodologies, detailing their assumptions, scope, and typical use cases, similarly to a family of software patterns~\cite{Gamma:1995:DPE}. This would assist practitioners in understanding the available options and making informed decisions based on their system’s characteristics and objectives. Additionally, establishing clear selection criteria could provide a structured way to evaluate methodologies, helping practitioners assess trade-offs between factors such as hardware efficiency, hyperparameter tuning, and overall system-wide energy consumption.

Another important consideration is the {\em interoperability between assessment methodologies}. Many existing approaches focus on isolated measurements, but there may be value in designing methodologies that can interface with one another, creating a more modular and adaptable assessment pipeline. This would facilitate comparisons across models, architectures, and computing environments, while also promoting a degree of standardization. Furthermore, as AI-enabled systems continue to evolve, mechanisms for methodology migration may become necessary to ensure assessments remain relevant and comparable over time. Developing protocols for transitioning between methodologies could help maintain continuity in energy assessments, even as best practices evolve.

At a broader level, the concept of a meta-methodology --a framework for guiding decision-making in energy assessment-- could be worth exploring. Previous work~\cite{DBLP:conf/icse/HaoLHG13,DBLP:journals/ese/OliveiraOCPF21} in the area of Software Engineering has leveraged this approach to help identify energy consumption hotspots in software systems. Rather than prescribing a single solution, a meta-methodology could help practitioners structure their approach based on the priorities relevant to their specific use case. For instance, it could offer guidance on whether capturing the energy cost of hyperparameter tuning is more relevant than measuring precise hardware consumption, depending on the context.
These ideas build on ongoing research into ML energy consumption. Existing studies have explored the energy costs of deep learning frameworks \cite{Alizadeh:2024:PES,Fontes:2024:SBU,georgiou2022green}, the impact of batching strategies on inference efficiency \cite{yarally2023batching}, and the role of quantization techniques in energy efficiency \cite{rajput2024benchmarking}. By integrating insights from this work into a more flexible and evolving framework, it may be possible to develop energy assessment methodologies that remain rigorous and adaptable as AI technologies continue to evolve.


\subsection{Positioning  Software Engineering}
Software engineering advances energy assessment in AI by enabling system-level and multi-granular measurement approaches that move beyond isolated model evaluation. 
We imagine that it will provide generalised structured yet flexible methodologies --meta-methodologies and adaptable guidelines--that support context-sensitive assessments across heterogeneous platforms. Additionally, Software Engineering approaches may contribute to the standardization and interoperability of tools and metrics, making energy measurements reproducible and comparable across systems and over time.

\section{Evaluating AI for Sustainability}


Assessing the energy consumption of AI-enabled systems requires {\em benchmarking frameworks} that incorporate both energy and carbon footprint metrics, alongside conventional measures such as accuracy, latency, and throughput. The objective is to develop methodologies that are widely adoptable and easily integrated into real-world CI/AI pipelines while being either hardware-aware or hardware-agnostic through normalization. These frameworks should account for both training and inference energy consumption, ensuring that energy assessments capture the full lifecycle of an AI model.

The need for such benchmarks originates from several pressing concerns. The environmental impact of AI is growing, with large-scale models consuming substantial energy and raising sustainability challenges (\cite{zhao2024end,bakhtiarifard2024ec}). All decision makers, such as developers, organizations, and policymakers, require reliable, meaningful, and comparable energy metrics (e.g., Joules, CO$_2$eq.) to make informed trade-offs between functional and extra-functional properties, such as accuracy, security, runtime, and energy cost. Additionally, potential regulatory requirements (e.g., the EU AI Act, ESG, and CSRD) may mandate energy reporting for AI-enabled systems. From an economic perspective, energy-efficient AI models offer business advantages, particularly in reducing operational costs in data centres and edge deployments. As AI regulation and standardization efforts evolve, energy measurement could become a critical service for evaluating AI models.

\subsection{Standardized Benchmarking for AI Sustainability}



A key step toward sustainable AI is the extension of existing benchmarking frameworks to include standardized energy and carbon metrics. Well-established benchmarking suites such as MLPerf \cite{reddi2020mlperf, mattson2020mlperf} or Hugging Face\footnote{\url{https://huggingface.co/}} leaderboards provide a structured foundation for measuring model performance, but currently with very limited standardized energy reporting. Extending these benchmarks to incorporate energy consumption across different layers of AI models, from individual operations to end-to-end pipelines, would enable a more comprehensive assessment framework.

Beyond energy measurement, integration with interdisciplinary research and policy standards is critical for ensuring broad adoption. Efforts such as the impact framework of the Green Software Foundation (.imp files) provide potential pathways for aligning AI benchmarking with emerging sustainability regulations. Additionally, defining a common ontology for AI energy benchmarking would facilitate interoperability across research domains and enable collaboration between computer scientists, policymakers, and industry stakeholders.

At the algorithmic level, energy-efficient neural architecture search (NAS) methods such as EC-NAS \cite{bakhtiarifard2024ec} and CE-NAS \cite{zhao2024end} have demonstrated the potential for optimizing architectures based on energy and carbon considerations. Similarly, approaches like Once-for-All (OFA) \cite{cai2019once}, which train a single network that can be specialized for different latency and energy constraints, illustrate the feasibility of integrating energy-awareness into AI model development.

\subsection{Multi-Dimensional Evaluation}

To ensure a holistic assessment of AI sustainability, energy and carbon footprint metrics must be evaluated alongside traditional performance measures such as accuracy and latency. A multi-dimensional evaluation framework would provide a clearer picture of trade-offs between computational cost and model effectiveness.\label{theme:holistic_evaluation}

A key aspect of this evaluation is the amortization of training energy costs over inference usage. Training large AI models is energy-intensive, but its impact on sustainability depends on how frequently the model is used. By treating training energy as a capital expense and distributing it across the total number of expected inferences, researchers can compute an amortized energy cost per inference, offering a more balanced sustainability metric.

Several existing frameworks already explore partial energy reporting. Hugging Face Model Cards sometimes include throughput or energy-related metrics, but these are not consistently standardized. Tools such as IrEne \cite{cao2021irene} and Smaragdine \cite{babakol2024tensor} provide entry points for AI energy accounting, but there is a need for a more structured methodology that aligns with AI benchmarking standards.\label{theme:standardization_benchmarking}

\subsection{Hardware and Normalized Metrics} \label{theme:metrics_hardware}
Ensuring that energy benchmarks are hardware-agnostic is essential for fair and meaningful comparisons across different AI models and deployment environments~\cite{bouza2023estimate}. However, hardware variations significantly impact energy consumption, making it challenging to compare models trained and deployed on different infrastructure.

One potential approach is to measure floating point operations (FLOPs) and integrate them into energy-prediction models, allowing researchers to estimate energy usage independently of specific hardware configurations. However, it is known that all FLOPs are not equal~\cite{10.1145/3545008.3545072}, and directly relying on FLOPs-driven energy measurements could bias these metrics~\cite{henderson2020towards}. Additionally, defining a baseline hardware profile could facilitate normalized energy reporting, ensuring that comparisons between models remain consistent regardless of hardware disparities.

Beyond FLOPs, reporting on broader resource utilization metrics (e.g., memory access patterns, I/O usage, and GPU/TPU power draw) could provide a more complete picture of energy consumption~\cite{anthony2020carbontracker}. A standardized methodology for energy-aware reporting across AI hardware platforms would help ensure that benchmarks remain interpretable and broadly applicable.  \label{theme:standardization_hardware}

Interdisciplinary collaborations between computer scientists, hardware engineers, and regulatory bodies could further strengthen AI benchmarking efforts by developing guidelines that integrate environmental sustainability concerns directly into AI deployment practices~\cite{wright2023efficiency}.

\subsection{Positioning  Software Engineering}
We imagine software engineering to advance Green AI by structuring how sustainability is evaluated and reported. Establishing and adopting standardized energy efficiency benchmarks may guide developers in evaluating, comparing, and optimizing the environmental impact of AI-enabled systems.
This includes integrating energy and carbon metrics into evaluation pipelines alongside accuracy and latency, applying normalization techniques to ensure fair cross-platform comparisons, and using lifecycle-aware models—such as amortized training cost—to contextualize energy consumption over time and usage scale.

\section{System Architecture and Life\-cy\-cle}

For AI-enabled systems, sustainability-driven architecting should integrate sustainability principles related to both lifecycle management (process) and architecture design (product). This means that sustainability-driven design decisions and quality assessment results should be considered throughout all phases of the lifecycle of AI-enabled software systems. Typical software architecture concepts~\cite{bass_software_2012} -- such as architectural tactics, scenarios, patterns, practices, indicators, metrics, and measures -- should be traceable to the corresponding AI-enabled architectural elements. Additionally, sustainability learning should provide guidance for dynamic adaptation in both MLOps and AI-enabled architectures.\label{theme:lifecycle_architecture}\label{theme:metrics_architecture}

\subsection{Architecting for Sustainable AI}
Current research on environmentally sustainable AI tends to emphasize improving the energy efficiency of individual models. However, AI-enabled systems often comprise multiple components -- both AI and non-AI ones -- that interact in complex ways. A narrow focus on isolated model performance may obscure the broader architecture context that ultimately determines system-wide sustainability~\cite{wright2023efficiency}. Adopting a system-level perspective enables sustainability considerations to be embedded throughout the AI-enabled architecture, from early design and deployment to runtime operation and long-term evolution.

%
%
From an architectural perspective, AI models function as specialized components that expose interfaces, interact with surrounding services, and influence the system's overall quality attributes~\cite{Lewis2021}. Established software architecture practices -- such as modular design, scenario-based evaluation, and trade-off analysis -- can guide the integration of AI components in ways that align with broader sustainability goals. Rather than selecting the most powerful model by default, architects should be able to consider which model best satisfies the system’s architecturally significant requirements~\cite{vallecillo_architectural_2022}, balancing accuracy, latency, and energy consumption.

%
%
Several design strategies can support these goals~\cite{jarvenpaa_synthesis_2024}. For example, selecting models that meet only the necessary performance requirements avoids energy waste due to overprovisioning~\cite{verdecchia_data-centric_2022}. Implementing lightweight models on the client side, with fallback to more complex models on the server, can improve efficiency in distributed deployments~\cite{xu_edgellm_2025}. Connector logic that dynamically routes requests based on resource constraints can further enhance adaptability~\cite{tedla_ecomls_2024,nijkamp_green_2024}. These decisions can be informed by predictive tools that estimate energy consumption at design time, helping architects evaluate tradeoffs early in the process.


In distributed AI-enabled systems, executing model workloads across edge, fog, and cloud layers enables more granular control of resource allocation and energy use. Refactoring legacy architectures using energy-aware tactics, and documenting architectural decisions alongside their energy implications, can strengthen traceability and operational transparency. Together, these practices enable the development of AI-enabled systems that prioritize sustainability without compromising performance, reliability, or maintainability across their lifecycle.


\subsection{Green AI at Runtime}
%
%
While training large AI models is energy-intensive, inference can account for a significant share of the system's total energy consumption, particularly when deployed at scale~\cite{Luccioni2024}. Addressing sustainability at runtime is therefore critical, yet remains less explored than energy-aware training practices. AI-enabled systems operate under diverse and often unpredictable runtime conditions, including changes in data quality, model performance, business needs, and infrastructure availability. These uncertainties impact both energy efficiency and quality of service.

%
To manage this complexity, runtime strategies must balance adaptability with sustainability. Enhancing observability -- the ability to monitor system behavior and internal states -- supports energy-aware decision-making by making deviations from expected performance and consumption patterns visible. In uncertain environments, observability enables early detection of inefficiencies and supports dynamic reconfiguration.

%
%
Self-adaptation mechanisms offer another path to improving runtime sustainability. By adjusting system structure or behavior in response to observed conditions, self-adaptive systems can optimize trade-offs between energy use and other quality attributes. Examples include dynamically selecting between models of varying complexity, reallocating workloads between edge and cloud resources, or retraining models on lower-energy hardware. These strategies are particularly valuable in MLOps pipelines, where frequent model updates and changing deployment contexts can otherwise lead to unnecessary energy costs.

Tools such as sustainability decision maps \cite{Lago2019-dm} can further support runtime reasoning by providing structured representations of trade-offs and system goals. These maps help track alignment between energy-related targets and system behavior over time, enabling more informed and transparent decisions. Integrating such tools into MLOps workflows can improve feedback loops and support continuous energy optimization~\cite{bhatt2024towards}.
%
Runtime sustainability requires more than localized model improvements --it calls for infrastructure-aware, adaptive, and transparent system-level coordination. Addressing energy use at this stage seems essential to ensuring that the long-term operation of AI-enabled systems remains aligned with sustainability goals.

\subsection{Systemic Sustainability in MLOps Workflows}
%
%

Sustainability efforts in AI development are often hindered by limited collaboration and information sharing across roles in the MLOps lifecycle. Although various metrics and logs are collected throughout model training, deployment, and monitoring, these data are rarely used holistically to inform sustainability-oriented decisions. Decisions made in isolation -- such as optimizing training energy while neglecting inference costs -- can lead to counterproductive outcomes when viewed from a system-wide perspective. \label{theme:lifecycle_mlops}

%
%
Improved traceability of energy-relevant decisions is essential to enable meaningful collaboration. Sustainability-related architectural elements -- such as design patterns, quality attributes, and runtime adaptations -- should be linked to observable data across the MLOps stack. Mechanisms for communication and alignment -- such as APIs that surface sustainability metrics between pipeline components -- can support consistent decision-making and make trade-offs across lifecycle phases more transparent.


A central challenge, however, is the lack of clearly defined responsibilities for sustainability. It is often unclear whether accountability lies with infrastructure providers, system architects, AI developers, or software engineers. This ambiguity prevents coordinated efforts to assess and improve sustainability across the full pipeline. Clarifying roles and responsibilities is a prerequisite for embedding sustainability into each stage of the development process.

Distinguishing between Green AI and Green SE may further support collaboration. While Green AI focuses on improving the energy efficiency of AI models themselves, Green SE emphasizes the sustainability of the full system in which those models are embedded. Clarifying this distinction can help assign accountability more effectively and ensure that sustainability is treated as a system-wide concern rather than the isolated optimization of a single model.



\subsection{Positioning Software Engineering}
We imagine software engineering to advance Green AI by treating AI as a software system component. By considering environmental sustainability across the entire life cycle including design and runtime, we adopt a holistic view on the sustainability impact of AI.

Software engineering approaches include modular design, component wise traceability, and runtime adaptability, which enable dynamic optimization of energy use across deployment contexts and division of responsibility. 

\section{Empirical Evaluation}

Research activity in Green AI is expanding rapidly, requiring a growing commitment to evidence-based approaches for enhancing the sustainability of AI-enabled systems. Empirical studies are increasingly used to investigate the environmental impacts of AI development and deployment~\cite{georgiou2022green,yarally2023batching,Luccioni2023,Luccioni2024,RajputTOSEM24,RajputMSR24,GongSSBSE24,aienergyscore-leaderboard} and to evaluate the level of improvement regarding sustainability of developed solutions achieved through newly proposed methods.

This growing body of empirical work has the potential to contribute to a shared knowledge base, enabling stakeholders to move beyond anecdotal evidence and toward informed, data-driven decision-making. Aligning methods and reporting practices allows better comparability between studies and facilitates the identification of robust patterns and best practices.

\subsection{Actionable Body of Knowledge}

Although empirical methods are becoming central to Green AI research, the results remain difficult to compare or generalize. This is largely due to methodological inconsistencies, varied study designs, and context-specific reporting. While methods such as controlled experiments and case studies are available, they are applied unevenly, limiting the ability to synthesize findings across studies.
This has led to a body of evidence that is difficult to accumulate or systematically synthesize.

A long-term vision for the field is to establish a coherent and evolving body of knowledge that allows stakeholders to make evidence-based decisions about sustainability practices in AI development and deployment. Such a shared foundation would enhance the comparability and generalizability of results and support the creation of benchmarks, inform regulatory efforts, and accelerate the transfer of knowledge from research to practice.

Realizing this vision requires coordinated efforts across the research community to align with empirical standards, promote replication, and foster shared methodological frameworks. Without such alignment, the field risks fragmentation and limited impact.
As Hart and Baehr~\cite{hart2013sustainable} argue, a sustainable body of knowledge in technical domains relies on both taxonomic structures and community engagement. The empirical Green AI community must thus evolve a structured landscape of open questions, replicated evidence, and reliable methodologies that can guide both researchers and practitioners.

\subsection{Standardized and Reliable Metrics}  \label{theme:standardized_metrics}

Sustainability assessments in software systems often rely on metrics that are inconsistently defined and unevenly applied, even when targeting comparable phenomena such as energy use or execution time. These inconsistencies emerge from a range of factors, including variation in hardware configurations, abstraction levels, and data collection tools~\cite{yarally2023batching,georgiou2022green}. Consequently, findings from different studies often resist meaningful comparison.\label{theme:metrics_taxonomy}

Such inconsistencies highlight a broader challenge: the lack of standardized and reliable metrics, which continues to obstruct efforts toward cumulative progress in Green AI research~\cite{anthony2020carbontracker,Luccioni2023}. Metrics used in this field span various abstraction levels, from low-level physical measures such as energy per instruction, to high-level proxies such as runtime or task throughput, and to more integrative compositions like sustainability scores \cite{selvan2025pepr,fatima2025} that, for instance, incorporate accuracy and explainability \cite{rajput2024benchmarking}. However, without consensus on their definitions, reliability, and the appropriate contexts for use, these metrics yield incompatible results that hinder replication and meta-analysis. Addressing this challenge would not only allow for systematic comparison between studies but would also provide a foundation for evidence-based guidelines and policy recommendations. It would support the identification of effective green practices and enable transparent and consistent reporting across academic and industrial settings~\cite{Luccioni2024}.

One possible direction is to develop a taxonomy of metrics that distinguishes between direct, proxy, and composite types. This taxonomy would ideally be supported by an articulation of each metric's assumptions, limitations, and the contexts in which they are most appropriate. Among composite metrics, a further distinction may be drawn between multi-dimensional and compositional approaches: while the former preserve trade-offs by reporting multiple values along separate axes (e.g., energy, latency, accuracy), the latter collapse these into a single score to support optimization or policy-facing decision-making. 
As understanding in the field evolves, so too might the criteria for empirical validation. There may also be value in exploring how standardized protocols for measurement can be developed and adapted over time -- particularly in methods such as case studies or surveys, where such practices are not yet established.  
Crucially, the standardization of metrics must be balanced with openness to innovation. New metrics may emerge in response to evolving technologies or new sustainability concerns. To manage this balance, a community-driven oversight mechanism -- perhaps in the form of a metrics committee -- could evaluate, endorse, and curate both existing and emerging measurement practices.

Lessons can be taken from the empirical software engineering community~\cite{Wohlin2024}. Further insights might be drawn from search-based software engineering research, where trade-off measurements and analysis have been well studied~\cite{10260877}, as well as from health research, where standardized data collection and reporting protocols have long provided a foundation for reproducibility and comparability~\cite{hart2013sustainable}.


\subsection{Positioning  Software Engineering}
We imagine software engineering to enhance Green AI by contributing empirical evaluation practices that enable trustworthy and generalizable sustainability claims. Drawing from empirical software engineering, we may adopt principles of methodological rigor, reproducibility, and structured comparison to support the development of a coherent evidence base. Software Engineering may also guide the creation of standardized metric taxonomies --distinguishing between direct, proxy, and composite measures-- and alignment of data protocols across tools and hardware. 
We envision that empirical research within Green AI and Green Software Engineering will continue to cross-fertilize, strengthening both fields.

\section{Education and Awareness}

Education and awareness play a crucial role in fostering sustainable AI practices. However, several challenges need to be addressed in order to effectively integrate 
Green AI principles into curricula and industry training programs.
Below, we highlight key challenges and considerations grouped by stage in the educational process.




Embedding sustainability into SE and AI education requires coordinated attention across several layers of the educational process~\cite{Peters+2024,10.1145/3708526}. At the curriculum level, integrating sustainability meaningfully requires institutional commitment and long-term thinking. While sustainability is inherently a long-term concern, academic structures often prioritize short-term outcomes. Institutions can leverage Intended Learning Outcomes (ILOs) to systematically integrate sustainability into educational goals, but must also be mindful of avoiding superficial virtue signalling. Gradual introduction of sustainability topics into existing, mandatory courses can increase reach beyond the self-selecting students typically drawn to electives.

Designing and implementing effective courses on sustainability-aware SE and AI poses practical barriers. Many Green AI courses rely on a wide array of prerequisites, spanning hardware, software, and systems-level topics. Hands-on activities are constrained by the need for specialized tools and equipment, and the lack of mature, accessible benchmarking frameworks. Furthermore, connecting energy consumption with more intuitive performance measures like time can be non-trivial. The design of assignments must also reflect the complexity of moving from energy assessment to actionable improvements, given the many interacting variables at play. To support effective learning, example systems used in teaching should be neither trivial nor overwhelming--ideally open-source and industry-relevant.

Students must navigate a high cognitive load when learning to develop sustainable software. The architectural concepts involved--such as patterns, tactics, styles, quality attributes, and trade-offs-- require deep understanding and differentiation. Thinking in terms of lifecycle sustainability adds further complexity, requiring students to reason across temporal and system boundaries. In addition, motivating students to value sustainability as a critical, multidimensional quality attribute is a challenge, particularly in industrial contexts where it is often not prioritized.

To assess the impact of sustainability teaching, educators must draw on robust pedagogical methods. Concepts like the Jevons Paradox, which highlights the counterintuitive effects of improved efficiency on total consumption, can be used to encourage critical thinking. Empirical evaluation approaches, such as A/B testing of teaching interventions, can help demonstrate effectiveness and inform iteration. Tying these evaluation methods to the intended learning outcomes supports the alignment between teaching goals and assessment practice. \label{theme:jevons_paradox}

These educational efforts are further complicated by the immaturity of the research field itself. In domains like Green AI, much of the work is mainly experimental and rapidly evolving. This makes it difficult to develop stable and reusable teaching packages. Moreover, translating low-level energy measurements into metrics that are meaningful for both technical and non-technical audiences remains an open challenge, requiring ongoing engagement from both educators and researchers.

\subsection{Aligning Software Engineering and AI}

We imagine that software engineering and AI education can both benefit from each other in our efforts to promote sustainability. 
While sustainability is not yet a core element in either field, SE’s system-level thinking and AI’s data-centric methods offer complementary strengths. By learning from each discipline’s approaches, we may grow educational practices that embed environmental awareness more deeply. We envision this exchange as a path toward a shared sustainability culture that strengthens both disciplines.

\section{Cross-Cutting Reflections}

Building on the synthesis provided in the discussion, this section identifies recurring themes that emerged across the workshop's diverse focus areas. These themes--\textit{standardization}, \textit{metrics}, and \textit{holistic and longitudinal thinking}-- cut across the topical boundaries of energy assessment, sustainability evaluation, architecture, empirical methodology, and education. They represent significant concerns for the development of environmentally sustainable AI.

\textbf{Standardization} was repeatedly highlighted as a fundamental need in measurement practices, benchmarking efforts, and empirical reporting. Discrepancies in the way energy or efficiency is defined (for example, joules, runtime, or CO$_2$) currently hinder comparability and prevent coordination between research, industry, and policy stakeholders. In the context of benchmarking frameworks, the lack of standardized reporting of energy and carbon consumption remains a major limitation (Sect. \ref{theme:standardization_benchmarking}). Similarly, proposals for normalized reporting across heterogeneous hardware platforms point to the necessity of standardization for interpretability and fairness (Sect. \ref{theme:standardization_hardware}). The development of a taxonomy of metrics, protocols for their application, and oversight mechanisms is central to advancing shared practices (Sect. \ref{theme:standardized_metrics}).

\textbf{Metrics} themselves constitute both a technical and conceptual challenge. As highlighted in the discussion on hardware-normalized metrics (Sect. \ref{theme:metrics_hardware}) and architectural decision-making (Sect. \ref{theme:metrics_architecture}), sustainability cannot be captured by a single scalar value. Multi-dimensional trade-offs, between accuracy, energy, latency, and explainability, require carefully chosen metrics. These must be context-sensitive yet comparable across cases. Empirical work has begun to surface classifications of direct, proxy, and composite metrics, but their assumptions and limitations remain under articulated (Sect. \ref{theme:metrics_taxonomy}).Developing metrics that are both robust and adaptable is essential for meaningful evaluation and progress.

\textbf{Holistic and longitudinal thinking} emerged as a critical counterbalance to narrow or isolated optimization efforts. Lifecycle sustainability --spanning design, training, deployment, and operation-- was emphasized in architecture discussions (Sect.~\ref{theme:lifecycle_architecture}) and sustainability evaluation frameworks (Sect. \ref{theme:holistic_evaluation}). However, workshop contributions also pointed to the importance of longer-term effects: how current benchmarking norms or energy optimizations may influence future systems and behaviors. The underutilization of longitudinal studies in empirical work 
and the potential for counterproductive outcomes such as the Jevons paradox (Sect.~\ref{theme:jevons_paradox}) underscore the need to assess not only a system's immediate footprint but its place within a larger temporal trajectory~\cite{wright2023efficiency}. Incorporating this perspective is particularly relevant for decision-making in MLOps contexts, where short-term gains may obscure longer-term sustainability costs (Sect. \ref{theme:lifecycle_mlops}).

Together, these themes suggest that advancing sustainable AI through software engineering requires more than isolated technical solutions. It demands a coordinated effort to establish shared foundations, articulate meaningful metrics, and reason across both system and temporal boundaries.

\section{Conclusion}
This article presented a synthesis of discussions from the “Greening AI with Software Engineering” workshop, which brought together researchers and practitioners to explore the environmental sustainability of AI through a software-centric lens. Rather than prescribing a single approach or framework, the contributions reflect a shared recognition that addressing AI’s environmental impact requires attention to multiple, interconnected dimensions.

The report is structured around five key areas: energy assessment and standardization, evaluation and benchmarking, architecture and lifecycle design, empirical methods and reproducibility, and education and awareness. Although each area raises distinct challenges, the discussions also revealed three cross-cutting themes--standardization, metrics, and holistic system thinking--that connect these areas and shape a broader research agenda.

What emerged is not a finished blueprint, but a conceptual research agenda: a set of aligned concerns, unresolved tensions, and promising directions. Moving forward, the research community is invited to build on these directions by developing shared benchmarks, improving observability and traceability, refining architectural practices, and embedding sustainability into both empirical studies and software engineering education. Addressing these challenges will require not only technical innovation, but also collaboration, openness, and long-term commitment from academia and industry. By articulating these challenges and points of convergence, this article contributes to an evolving foundation for the development of environmentally sustainable AI with software engineering.

\section{Acknowledgement}
We would like to thank the \textit{Centre Européen de Calcul Atomique et Moléculaire} (\href{https://www.cecam.org/}{CECAM}) and the \href{https://www.lorentzcenter.nl/}{Lorentz Center} for their supporting the  “Greening AI with Software Engineering” workshop, which served as the foundation for the collaborative research summarized in this report.

\bibliographystyle{ACM-Reference-Format} 
\bibliography{biblio}


\begin{thebibliography}{48}


\ifx \showCODEN    \undefined \def \showCODEN     #1{\unskip}     \fi
\ifx \showISBNx    \undefined \def \showISBNx     #1{\unskip}     \fi
\ifx \showISBNxiii \undefined \def \showISBNxiii  #1{\unskip}     \fi
\ifx \showISSN     \undefined \def \showISSN      #1{\unskip}     \fi
\ifx \showLCCN     \undefined \def \showLCCN      #1{\unskip}     \fi
\ifx \shownote     \undefined \def \shownote      #1{#1}          \fi
\ifx \showarticletitle \undefined \def \showarticletitle #1{#1}   \fi
\ifx \showURL      \undefined \def \showURL       {\relax}        \fi
\providecommand\bibfield[2]{#2}
\providecommand\bibinfo[2]{#2}
\providecommand\natexlab[1]{#1}
\providecommand\showeprint[2][]{arXiv:#2}

\bibitem[Alizadeh et~al\mbox{.}(2025)]%
        {alizadeh2024language}
\bibfield{author}{\bibinfo{person}{Negar Alizadeh}, \bibinfo{person}{Boris
  Belchev}, \bibinfo{person}{Nishant Saurabh}, \bibinfo{person}{Patricia
  Kelbert}, {and} \bibinfo{person}{Fernando Castor}.}
  \bibinfo{year}{2025}\natexlab{}.
\newblock \showarticletitle{Language Models in Software Development Tasks: An
  Experimental Analysis of Energy and Accuracy}.
\newblock  (\bibinfo{year}{2025}).
\newblock
\showeprint[arxiv]{2412.00329}
\urldef\tempurl%
\url{https://arxiv.org/abs/2412.00329}
\showURL{%
\tempurl}
\newblock
\shownote{To appear in the Proceedings of the 22nd International Conference on
  Mining Software Repositories, {MSR} 2025, Ottawa, Canada, April 28-29, 2025}.


\bibitem[Alizadeh and Castor(2024)]%
        {Alizadeh:2024:PES}
\bibfield{author}{\bibinfo{person}{Negar Alizadeh} {and}
  \bibinfo{person}{Fernando Castor}.} \bibinfo{year}{2024}\natexlab{}.
\newblock \showarticletitle{Green AI: a Preliminary Empirical Study on Energy
  Consumption in DL Models Across Different Runtime Infrastructures}. In
  \bibinfo{booktitle}{\emph{Proceedings of the IEEE/ACM 3rd International
  Conference on AI Engineering - Software Engineering for AI}} (Lisbon,
  Portugal) \emph{(\bibinfo{series}{CAIN '24})}. \bibinfo{publisher}{{ACM}},
  \bibinfo{address}{New York, NY, USA}, \bibinfo{pages}{134–139}.
\newblock
\showISBNx{9798400705915}
\href{https://doi.org/10.1145/3644815.3644967}{doi:\nolinkurl{10.1145/3644815.3644967}}


\bibitem[Anthony et~al\mbox{.}(2020)]%
        {anthony2020carbontracker}
\bibfield{author}{\bibinfo{person}{Lasse F.~Wolff Anthony},
  \bibinfo{person}{Benjamin Kanding}, {and} \bibinfo{person}{Raghavendra
  Selvan}.} \bibinfo{year}{2020}\natexlab{}.
\newblock \showarticletitle{Carbontracker: Tracking and Predicting the Carbon
  Footprint of Training Deep Learning Models}.
\newblock  (\bibinfo{year}{2020}).
\newblock
\showeprint[arxiv]{2007.03051}~[cs.CY]
\urldef\tempurl%
\url{https://arxiv.org/abs/2007.03051}
\showURL{%
\tempurl}
\newblock
\shownote{Presented at ICML Workshop on Challenges in Deploying and Monitoring
  Machine Learning Systems}.


\bibitem[Babakol and Liu(2024)]%
        {babakol2024tensor}
\bibfield{author}{\bibinfo{person}{Timur Babakol} {and}
  \bibinfo{person}{Yu~David Liu}.} \bibinfo{year}{2024}\natexlab{}.
\newblock \showarticletitle{Tensor-Aware Energy Accounting}. In
  \bibinfo{booktitle}{\emph{Proceedings of the IEEE/ACM 46th International
  Conference on Software Engineering}} (Lisbon, Portugal)
  \emph{(\bibinfo{series}{ICSE '24})}. \bibinfo{publisher}{{ACM}},
  \bibinfo{address}{New York, NY, USA}, Article \bibinfo{articleno}{93},
  \bibinfo{numpages}{12}~pages.
\newblock
\showISBNx{9798400702174}
\href{https://doi.org/10.1145/3597503.3639156}{doi:\nolinkurl{10.1145/3597503.3639156}}


\bibitem[Bakhtiarifard et~al\mbox{.}(2024)]%
        {bakhtiarifard2024ec}
\bibfield{author}{\bibinfo{person}{Pedram Bakhtiarifard},
  \bibinfo{person}{Christian Igel}, {and} \bibinfo{person}{Raghavendra
  Selvan}.} \bibinfo{year}{2024}\natexlab{}.
\newblock \showarticletitle{EC-NAS: Energy Consumption Aware Tabular Benchmarks
  for Neural Architecture Search}. In \bibinfo{booktitle}{\emph{{IEEE}
  International Conference on Acoustics, Speech and Signal Processing, {ICASSP}
  2024, Seoul, Republic of Korea, April 14-19, 2024}}.
  \bibinfo{publisher}{{IEEE}}, \bibinfo{pages}{5660--5664}.
\newblock
\href{https://doi.org/10.1109/ICASSP48485.2024.10448303}{doi:\nolinkurl{10.1109/ICASSP48485.2024.10448303}}


\bibitem[Bass et~al\mbox{.}(2012)]%
        {bass_software_2012}
\bibfield{author}{\bibinfo{person}{Len Bass}, \bibinfo{person}{Paul Clements},
  {and} \bibinfo{person}{Rick Kazman}.} \bibinfo{year}{2012}\natexlab{}.
\newblock \bibinfo{booktitle}{\emph{Software {Architecture} in {Practice}}
  (\bibinfo{edition}{3rd} ed.)}.
\newblock \bibinfo{publisher}{Addison-Wesley Professional},
  \bibinfo{address}{Westford, MA, USA}.
\newblock
\showISBNx{978-0-321-81573-6}
\newblock
\shownote{Publication Title: The SEI Series in Software Engineering}.


\bibitem[Bhatt et~al\mbox{.}(2024)]%
        {bhatt2024towards}
\bibfield{author}{\bibinfo{person}{Hiya Bhatt}, \bibinfo{person}{Shrikara
  Arun}, \bibinfo{person}{Adyansh Kakran}, {and} \bibinfo{person}{Karthik
  Vaidhyanathan}.} \bibinfo{year}{2024}\natexlab{}.
\newblock \showarticletitle{{ Towards Architecting Sustainable MLOps: A
  Self-Adaptation Approach }}. In \bibinfo{booktitle}{\emph{2024 IEEE 21st
  International Conference on Software Architecture Companion (ICSA-C)}}.
  \bibinfo{publisher}{{IEEE}}, \bibinfo{address}{Los Alamitos, CA, USA},
  \bibinfo{pages}{179--182}.
\newblock
\href{https://doi.org/10.1109/ICSA-C63560.2024.00038}{doi:\nolinkurl{10.1109/ICSA-C63560.2024.00038}}


\bibitem[Bouza et~al\mbox{.}(2023)]%
        {bouza2023estimate}
\bibfield{author}{\bibinfo{person}{Luc{\'\i}a Bouza},
  \bibinfo{person}{Aur{\'e}lie Bugeau}, {and} \bibinfo{person}{Lo{\"\i}c
  Lannelongue}.} \bibinfo{year}{2023}\natexlab{}.
\newblock \showarticletitle{How to estimate carbon footprint when training deep
  learning models? A guide and review}.
\newblock \bibinfo{journal}{\emph{Environmental Research Communications}}
  \bibinfo{volume}{5}, \bibinfo{number}{11} (\bibinfo{date}{nov}
  \bibinfo{year}{2023}), \bibinfo{pages}{115014}.
\newblock
\href{https://doi.org/10.1088/2515-7620/acf81b}{doi:\nolinkurl{10.1088/2515-7620/acf81b}}


\bibitem[Cai et~al\mbox{.}(2020)]%
        {cai2019once}
\bibfield{author}{\bibinfo{person}{Han Cai}, \bibinfo{person}{Chuang Gan},
  \bibinfo{person}{Tianzhe Wang}, \bibinfo{person}{Zhekai Zhang}, {and}
  \bibinfo{person}{Song Han}.} \bibinfo{year}{2020}\natexlab{}.
\newblock \showarticletitle{Once-for-All: Train One Network and Specialize it
  for Efficient Deployment}.
\newblock  (\bibinfo{year}{2020}).
\newblock
\showeprint[arxiv]{1908.09791}~[cs.LG]
\urldef\tempurl%
\url{https://arxiv.org/abs/1908.09791}
\showURL{%
\tempurl}


\bibitem[Cao et~al\mbox{.}(2021)]%
        {cao2021irene}
\bibfield{author}{\bibinfo{person}{Qingqing Cao}, \bibinfo{person}{Yash~Kumar
  Lal}, \bibinfo{person}{Harsh Trivedi}, \bibinfo{person}{Aruna
  Balasubramanian}, {and} \bibinfo{person}{Niranjan Balasubramanian}.}
  \bibinfo{year}{2021}\natexlab{}.
\newblock \showarticletitle{IrEne: Interpretable Energy Prediction for
  Transformers}.
\newblock  (\bibinfo{year}{2021}).
\newblock
\showeprint[arxiv]{2106.01199}~[cs.CL]
\urldef\tempurl%
\url{https://arxiv.org/abs/2106.01199}
\showURL{%
\tempurl}


\bibitem[Fatima et~al\mbox{.}(2025)]%
        {fatima2025}
\bibfield{author}{\bibinfo{person}{Iffat Fatima}, \bibinfo{person}{Patricia
  Lago}, \bibinfo{person}{Vasilios Andrikopoulos}, {and} \bibinfo{person}{Bram
  van~der Waaij}.} \bibinfo{year}{2025}\natexlab{}.
\newblock \showarticletitle{Using Sustainability Impact Scores for Software
  Architecture Evaluation}.
\newblock  (\bibinfo{year}{2025}).
\newblock
\showeprint[arxiv]{2501.17004}~[cs.SE]
\urldef\tempurl%
\url{https://arxiv.org/abs/2501.17004}
\showURL{%
\tempurl}
\newblock
\shownote{{To appear in the Proceedings of the 22nd International Conference on
  Software Architecture (ICSA)}}.


\bibitem[Forti et~al\mbox{.}(2022)]%
        {SEN_aim}
\bibfield{author}{\bibinfo{person}{Stefano Forti}, \bibinfo{person}{Uwe
  Breitenb\"{u}cher}, {and} \bibinfo{person}{Jacopo Soldani}.}
  \bibinfo{year}{2022}\natexlab{}.
\newblock \showarticletitle{Trending Topics in Software Engineering}.
\newblock \bibinfo{journal}{\emph{SIGSOFT Softw. Eng. Notes}}
  \bibinfo{volume}{47}, \bibinfo{number}{3} (\bibinfo{date}{July}
  \bibinfo{year}{2022}), \bibinfo{pages}{20–21}.
\newblock
\showISSN{0163-5948}
\href{https://doi.org/10.1145/3539814.3539820}{doi:\nolinkurl{10.1145/3539814.3539820}}


\bibitem[Franch et~al\mbox{.}(2022)]%
        {vallecillo_architectural_2022}
\bibfield{author}{\bibinfo{person}{Xavier Franch}, \bibinfo{person}{Silverio
  Martínez-Fernández}, \bibinfo{person}{Claudia~P. Ayala}, {and}
  \bibinfo{person}{Cristina Gómez}.} \bibinfo{year}{2022}\natexlab{}.
\newblock \showarticletitle{Architectural {Decisions} in {AI}-{Based}
  {Systems}: {An} {Ontological} {View}}.
\newblock In \bibinfo{booktitle}{\emph{Quality of {Information} and
  {Communications} {Technology}}}, \bibfield{editor}{\bibinfo{person}{Antonio
  Vallecillo}, \bibinfo{person}{Joost Visser}, {and} \bibinfo{person}{Ricardo
  Pérez-Castillo}} (Eds.). Vol.~\bibinfo{volume}{1621}.
  \bibinfo{publisher}{Springer International Publishing},
  \bibinfo{address}{Cham}, \bibinfo{pages}{18--27}.
\newblock
\showISBNx{978-3-031-14178-2 978-3-031-14179-9}
\href{https://doi.org/10.1007/978-3-031-14179-9_2}{doi:\nolinkurl{10.1007/978-3-031-14179-9_2}}
\newblock
\shownote{Series Title: Communications in Computer and Information Science}.


\bibitem[Gamma et~al\mbox{.}(1995)]%
        {Gamma:1995:DPE}
\bibfield{author}{\bibinfo{person}{Erich Gamma}, \bibinfo{person}{Richard
  Helm}, \bibinfo{person}{Ralph Johnson}, {and} \bibinfo{person}{John
  Vlissides}.} \bibinfo{year}{1995}\natexlab{}.
\newblock \bibinfo{booktitle}{\emph{Design patterns: elements of reusable
  object-oriented software}}.
\newblock \bibinfo{publisher}{Addison-Wesley Longman Publishing Co., Inc.},
  \bibinfo{address}{USA}.
\newblock
\showISBNx{0201633612}


\bibitem[Georgiou et~al\mbox{.}(2022)]%
        {georgiou2022green}
\bibfield{author}{\bibinfo{person}{Stefanos Georgiou}, \bibinfo{person}{Maria
  Kechagia}, \bibinfo{person}{Tushar Sharma}, \bibinfo{person}{Federica Sarro},
  {and} \bibinfo{person}{Ying Zou}.} \bibinfo{year}{2022}\natexlab{}.
\newblock \showarticletitle{Green AI: do deep learning frameworks have
  different costs?}. In \bibinfo{booktitle}{\emph{Proceedings of the 44th
  International Conference on Software Engineering}} (Pittsburgh, Pennsylvania)
  \emph{(\bibinfo{series}{ICSE '22})}. \bibinfo{publisher}{{ACM}},
  \bibinfo{address}{New York, NY, USA}, \bibinfo{pages}{1082–1094}.
\newblock
\showISBNx{9781450392211}
\href{https://doi.org/10.1145/3510003.3510221}{doi:\nolinkurl{10.1145/3510003.3510221}}


\bibitem[Gong et~al\mbox{.}(2024)]%
        {GongSSBSE24}
\bibfield{author}{\bibinfo{person}{Jingzhi Gong}, \bibinfo{person}{Sisi Li},
  \bibinfo{person}{Giordano d'Aloisio}, \bibinfo{person}{Zishuo Ding},
  \bibinfo{person}{Yulong Ye}, \bibinfo{person}{William~B. Langdon}, {and}
  \bibinfo{person}{Federica Sarro}.} \bibinfo{year}{2024}\natexlab{}.
\newblock \showarticletitle{GreenStableYolo: Optimizing Inference Time and
  Image Quality of Text-to-Image Generation}. In
  \bibinfo{booktitle}{\emph{Search-Based Software Engineering}},
  \bibfield{editor}{\bibinfo{person}{Gunel Jahangirova} {and}
  \bibinfo{person}{Foutse Khomh}} (Eds.). \bibinfo{publisher}{Springer Nature
  Switzerland}, \bibinfo{address}{Cham}, \bibinfo{pages}{70--76}.
\newblock
\showISBNx{978-3-031-64573-0}


\bibitem[Hao et~al\mbox{.}(2013)]%
        {DBLP:conf/icse/HaoLHG13}
\bibfield{author}{\bibinfo{person}{Shuai Hao}, \bibinfo{person}{Ding Li},
  \bibinfo{person}{William G.~J. Halfond}, {and} \bibinfo{person}{Ramesh
  Govindan}.} \bibinfo{year}{2013}\natexlab{}.
\newblock \showarticletitle{{ Estimating mobile application energy consumption
  using program analysis }}. In \bibinfo{booktitle}{\emph{2013 35th
  International Conference on Software Engineering (ICSE)}}.
  \bibinfo{publisher}{{IEEE}}, \bibinfo{address}{Los Alamitos, CA, USA},
  \bibinfo{pages}{92--101}.
\newblock
\href{https://doi.org/10.1109/ICSE.2013.6606555}{doi:\nolinkurl{10.1109/ICSE.2013.6606555}}


\bibitem[Hart and Baehr(2013)]%
        {hart2013sustainable}
\bibfield{author}{\bibinfo{person}{Hillary Hart} {and} \bibinfo{person}{Craig
  Baehr}.} \bibinfo{year}{2013}\natexlab{}.
\newblock \showarticletitle{Sustainable Practices for Developing a Body of
  Knowledge}.
\newblock \bibinfo{journal}{\emph{Technical Communication}}
  \bibinfo{volume}{60}, \bibinfo{number}{4} (\bibinfo{date}{November}
  \bibinfo{year}{2013}), \bibinfo{pages}{259--266}.
\newblock
\urldef\tempurl%
\url{https://www.jstor.org/stable/26464355}
\showURL{%
\tempurl}


\bibitem[Henderson et~al\mbox{.}(2020)]%
        {henderson2020towards}
\bibfield{author}{\bibinfo{person}{Peter Henderson}, \bibinfo{person}{Jieru
  Hu}, \bibinfo{person}{Joshua Romoff}, \bibinfo{person}{Emma Brunskill},
  \bibinfo{person}{Dan Jurafsky}, {and} \bibinfo{person}{Joelle Pineau}.}
  \bibinfo{year}{2020}\natexlab{}.
\newblock \showarticletitle{Towards the systematic reporting of the energy and
  carbon footprints of machine learning}.
\newblock \bibinfo{journal}{\emph{Journal of Machine Learning Research}}
  \bibinfo{volume}{21}, \bibinfo{number}{248} (\bibinfo{year}{2020}),
  \bibinfo{pages}{1--43}.
\newblock


\bibitem[Jacques et~al\mbox{.}(2024)]%
        {Fontes:2024:SBU}
\bibfield{author}{\bibinfo{person}{Vitor Maciel~Fontes Jacques},
  \bibinfo{person}{Negar Alizadeh}, {and} \bibinfo{person}{Fernando Castor}.}
  \bibinfo{year}{2024}\natexlab{}.
\newblock \showarticletitle{A Study on the Battery Usage of Deep Learning
  Frameworks on iOS Devices}. In \bibinfo{booktitle}{\emph{Proceedings of the
  IEEE/ACM 11th International Conference on Mobile Software Engineering and
  Systems}} (Lisbon, Portugal) \emph{(\bibinfo{series}{MOBILESoft '24})}.
  \bibinfo{publisher}{{ACM}}, \bibinfo{address}{New York, NY, USA},
  \bibinfo{pages}{1–11}.
\newblock
\showISBNx{9798400705946}
\href{https://doi.org/10.1145/3647632.3647990}{doi:\nolinkurl{10.1145/3647632.3647990}}


\bibitem[Järvenpää et~al\mbox{.}(2024)]%
        {jarvenpaa_synthesis_2024}
\bibfield{author}{\bibinfo{person}{Heli Järvenpää},
  \bibinfo{person}{Patricia Lago}, \bibinfo{person}{Justus Bogner},
  \bibinfo{person}{Grace Lewis}, \bibinfo{person}{Henry Muccini}, {and}
  \bibinfo{person}{Ipek Ozkaya}.} \bibinfo{year}{2024}\natexlab{}.
\newblock \showarticletitle{A {Synthesis} of {Green} {Architectural} {Tactics}
  for {ML}-{Enabled} {Systems}}. In \bibinfo{booktitle}{\emph{Proceedings of
  the 46th {International} {Conference} on {Software} {Engineering}: {Software}
  {Engineering} in {Society}}}. \bibinfo{publisher}{ACM},
  \bibinfo{address}{Lisbon Portugal}, \bibinfo{pages}{130--141}.
\newblock
\showISBNx{9798400704994}
\href{https://doi.org/10.1145/3639475.3640111}{doi:\nolinkurl{10.1145/3639475.3640111}}


\bibitem[Lago(2019)]%
        {Lago2019-dm}
\bibfield{author}{\bibinfo{person}{Patricia Lago}.}
  \bibinfo{year}{2019}\natexlab{}.
\newblock \showarticletitle{{ Architecture Design Decision Maps for Software
  Sustainability }}. In \bibinfo{booktitle}{\emph{2019 IEEE/ACM 41st
  International Conference on Software Engineering: Software Engineering in
  Society (ICSE-SEIS)}}. \bibinfo{publisher}{{IEEE}}, \bibinfo{address}{Los
  Alamitos, CA, USA}, \bibinfo{pages}{61--64}.
\newblock
\href{https://doi.org/10.1109/ICSE-SEIS.2019.00015}{doi:\nolinkurl{10.1109/ICSE-SEIS.2019.00015}}


\bibitem[Lewis et~al\mbox{.}(2021)]%
        {Lewis2021}
\bibfield{author}{\bibinfo{person}{Grace~A. Lewis}, \bibinfo{person}{Ipek
  Ozkaya}, {and} \bibinfo{person}{Xiwei Xu}.} \bibinfo{year}{2021}\natexlab{}.
\newblock \showarticletitle{{ Software Architecture Challenges for ML Systems
  }}. In \bibinfo{booktitle}{\emph{2021 IEEE International Conference on
  Software Maintenance and Evolution (ICSME)}}. \bibinfo{publisher}{{IEEE}},
  \bibinfo{address}{Los Alamitos, CA, USA}, \bibinfo{pages}{634--638}.
\newblock
\href{https://doi.org/10.1109/ICSME52107.2021.00071}{doi:\nolinkurl{10.1109/ICSME52107.2021.00071}}


\bibitem[L\'{o}pez et~al\mbox{.}(2023)]%
        {10.1145/3545008.3545072}
\bibfield{author}{\bibinfo{person}{Francisco L\'{o}pez}, \bibinfo{person}{Lars
  Karlsson}, {and} \bibinfo{person}{Paolo Bientinesi}.}
  \bibinfo{year}{2023}\natexlab{}.
\newblock \showarticletitle{FLOPs as a Discriminant for Dense Linear Algebra
  Algorithms}. In \bibinfo{booktitle}{\emph{Proceedings of the 51st
  International Conference on Parallel Processing}} (Bordeaux, France)
  \emph{(\bibinfo{series}{ICPP '22})}. \bibinfo{publisher}{{ACM}},
  \bibinfo{address}{New York, NY, USA}, Article \bibinfo{articleno}{11},
  \bibinfo{numpages}{10}~pages.
\newblock
\showISBNx{9781450397339}
\href{https://doi.org/10.1145/3545008.3545072}{doi:\nolinkurl{10.1145/3545008.3545072}}


\bibitem[Luccioni et~al\mbox{.}(2023)]%
        {Luccioni2023}
\bibfield{author}{\bibinfo{person}{Alexandra~Sasha Luccioni},
  \bibinfo{person}{Sylvain Viguier}, {and} \bibinfo{person}{Anne-Laure
  Ligozat}.} \bibinfo{year}{2023}\natexlab{}.
\newblock \showarticletitle{Estimating the Carbon Footprint of BLOOM, a 176B
  Parameter Language Model}.
\newblock \bibinfo{journal}{\emph{Journal of Machine Learning Research}}
  \bibinfo{volume}{24}, \bibinfo{number}{253} (\bibinfo{year}{2023}),
  \bibinfo{pages}{1--15}.
\newblock
\urldef\tempurl%
\url{http://jmlr.org/papers/v24/23-0069.html}
\showURL{%
\tempurl}


\bibitem[Luccioni et~al\mbox{.}(2025)]%
        {aienergyscore-leaderboard}
\bibfield{author}{\bibinfo{person}{Sasha Luccioni}, \bibinfo{person}{Boris
  Gamazaychikov}, \bibinfo{person}{Emma Strubell}, \bibinfo{person}{Sara
  Hooker}, \bibinfo{person}{Yacine Jernite}, \bibinfo{person}{Carole-Jean Wu},
  {and} \bibinfo{person}{Margaret Mitchell}.} \bibinfo{year}{2025}\natexlab{}.
\newblock \bibinfo{title}{AI Energy Score Leaderboard - February 2025}.
\newblock
\urldef\tempurl%
\url{https://huggingface.co/spaces/AIEnergyScore/Leaderboard}
\showURL{%
\tempurl}


\bibitem[Luccioni et~al\mbox{.}(2024)]%
        {Luccioni2024}
\bibfield{author}{\bibinfo{person}{Sasha Luccioni}, \bibinfo{person}{Yacine
  Jernite}, {and} \bibinfo{person}{Emma Strubell}.}
  \bibinfo{year}{2024}\natexlab{}.
\newblock \showarticletitle{{Power Hungry Processing: Watts Driving the Cost of
  AI Deployment?}}. In \bibinfo{booktitle}{\emph{Proceedings of the 2024 ACM
  Conference on Fairness, Accountability, and Transparency}} (Rio de Janeiro,
  Brazil) \emph{(\bibinfo{series}{FAccT '24})}. \bibinfo{publisher}{{ACM}},
  \bibinfo{address}{New York, NY, USA}, \bibinfo{pages}{85–99}.
\newblock
\showISBNx{9798400704505}
\href{https://doi.org/10.1145/3630106.3658542}{doi:\nolinkurl{10.1145/3630106.3658542}}


\bibitem[Mattson et~al\mbox{.}(2020)]%
        {mattson2020mlperf}
\bibfield{author}{\bibinfo{person}{Peter Mattson}, \bibinfo{person}{Christine
  Cheng}, \bibinfo{person}{Gregory~F. Diamos}, \bibinfo{person}{Cody Coleman},
  \bibinfo{person}{Paulius Micikevicius}, \bibinfo{person}{David~A. Patterson},
  \bibinfo{person}{Hanlin Tang}, \bibinfo{person}{Gu{-}Yeon Wei},
  \bibinfo{person}{Peter Bailis}, \bibinfo{person}{Victor Bittorf},
  \bibinfo{person}{David Brooks}, \bibinfo{person}{Dehao Chen},
  \bibinfo{person}{Debo Dutta}, \bibinfo{person}{Udit Gupta},
  \bibinfo{person}{Kim~M. Hazelwood}, \bibinfo{person}{Andy Hock},
  \bibinfo{person}{Xinyuan Huang}, \bibinfo{person}{Daniel Kang},
  \bibinfo{person}{David Kanter}, \bibinfo{person}{Naveen Kumar},
  \bibinfo{person}{Jeffery Liao}, \bibinfo{person}{Deepak Narayanan},
  \bibinfo{person}{Tayo Oguntebi}, \bibinfo{person}{Gennady Pekhimenko},
  \bibinfo{person}{Lillian Pentecost}, \bibinfo{person}{Vijay~Janapa Reddi},
  \bibinfo{person}{Taylor Robie}, \bibinfo{person}{Tom~St. John},
  \bibinfo{person}{Carole{-}Jean Wu}, \bibinfo{person}{Lingjie Xu},
  \bibinfo{person}{Cliff Young}, {and} \bibinfo{person}{Matei Zaharia}.}
  \bibinfo{year}{2020}\natexlab{}.
\newblock \showarticletitle{MLPerf Training Benchmark}. In
  \bibinfo{booktitle}{\emph{Proceedings of the Third Conference on Machine
  Learning and Systems, MLSys 2020, Austin, TX, USA, March 2-4, 2020}},
  \bibfield{editor}{\bibinfo{person}{Inderjit~S. Dhillon},
  \bibinfo{person}{Dimitris~S. Papailiopoulos}, {and} \bibinfo{person}{Vivienne
  Sze}} (Eds.). \bibinfo{publisher}{mlsys.org}.
\newblock
\urldef\tempurl%
\url{https://proceedings.mlsys.org/paper\_files/paper/2020/hash/411e39b117e885341f25efb8912945f7-Abstract.html}
\showURL{%
\tempurl}


\bibitem[Menzies and Johnson(2024)]%
        {MenziesIEEEsw24}
\bibfield{author}{\bibinfo{person}{Tim Menzies} {and} \bibinfo{person}{Brittany
  Johnson}.} \bibinfo{year}{2024}\natexlab{}.
\newblock \showarticletitle{{ Powering Down: An Interview With Federica Sarro
  on Tackling Energy Consumption in AI-Powered Software Systems }}.
\newblock \bibinfo{journal}{\emph{IEEE Software}} \bibinfo{volume}{41},
  \bibinfo{number}{05} (\bibinfo{date}{Sept.} \bibinfo{year}{2024}),
  \bibinfo{pages}{89--92}.
\newblock
\showISSN{1937-4194}
\href{https://doi.org/10.1109/MS.2024.3410011}{doi:\nolinkurl{10.1109/MS.2024.3410011}}


\bibitem[Moreira et~al\mbox{.}(2025)]%
        {10.1145/3708526}
\bibfield{author}{\bibinfo{person}{Ana Moreira}, \bibinfo{person}{Patricia
  Lago}, \bibinfo{person}{Rogardt Heldal}, \bibinfo{person}{Stefanie Betz},
  \bibinfo{person}{Ian Brooks}, \bibinfo{person}{Rafael Capilla},
  \bibinfo{person}{Vlad~Constantin Coroam\u{a}}, \bibinfo{person}{Leticia
  Duboc}, \bibinfo{person}{Jo\~{a}o~Paulo Fernandes}, \bibinfo{person}{Ola
  Leifler}, \bibinfo{person}{Ngoc-Thanh Nguyen}, \bibinfo{person}{Shola
  Oyedeji}, \bibinfo{person}{Birgit Penzenstadler},
  \bibinfo{person}{Anne-Kathrin Peters}, \bibinfo{person}{Jari Porras}, {and}
  \bibinfo{person}{Colin~C. Venters}.} \bibinfo{year}{2025}\natexlab{}.
\newblock \showarticletitle{A Roadmap for Integrating Sustainability into
  Software Engineering Education}.
\newblock \bibinfo{journal}{\emph{ACM Trans. Softw. Eng. Methodol.}}
  \bibinfo{volume}{34}, \bibinfo{number}{5}, Article \bibinfo{articleno}{139}
  (\bibinfo{date}{May} \bibinfo{year}{2025}), \bibinfo{numpages}{27}~pages.
\newblock
\showISSN{1049-331X}
\href{https://doi.org/10.1145/3708526}{doi:\nolinkurl{10.1145/3708526}}


\bibitem[Nijkamp et~al\mbox{.}(2024)]%
        {nijkamp_green_2024}
\bibfield{author}{\bibinfo{person}{Nienke Nijkamp}, \bibinfo{person}{June
  Sallou}, \bibinfo{person}{Niels Van Der~Heijden}, {and}
  \bibinfo{person}{Luís Cruz}.} \bibinfo{year}{2024}\natexlab{}.
\newblock \showarticletitle{Green {AI} in {Action}: {Strategic} {Model}
  {Selection} for {Ensembles} in {Production}}. In
  \bibinfo{booktitle}{\emph{Proceedings of the 1st {ACM} {International}
  {Conference} on {AI}-{Powered} {Software}}}. \bibinfo{publisher}{ACM},
  \bibinfo{address}{Porto de Galinhas Brazil}, \bibinfo{pages}{50--58}.
\newblock
\showISBNx{9798400706851}
\href{https://doi.org/10.1145/3664646.3664763}{doi:\nolinkurl{10.1145/3664646.3664763}}


\bibitem[Oliveira et~al\mbox{.}(2021)]%
        {DBLP:journals/ese/OliveiraOCPF21}
\bibfield{author}{\bibinfo{person}{Wellington Oliveira},
  \bibinfo{person}{Renato Oliveira}, \bibinfo{person}{Fernando Castor},
  \bibinfo{person}{Gustavo Pinto}, {and} \bibinfo{person}{Jo\~{a}o~Paulo
  Fernandes}.} \bibinfo{year}{2021}\natexlab{}.
\newblock \showarticletitle{Improving energy-efficiency by recommending Java
  collections}.
\newblock \bibinfo{journal}{\emph{Empirical Softw. Engg.}}
  \bibinfo{volume}{26}, \bibinfo{number}{3} (\bibinfo{date}{May}
  \bibinfo{year}{2021}), \bibinfo{numpages}{45}~pages.
\newblock
\showISSN{1382-3256}
\href{https://doi.org/10.1007/s10664-021-09950-y}{doi:\nolinkurl{10.1007/s10664-021-09950-y}}


\bibitem[Peters et~al\mbox{.}(2024)]%
        {Peters+2024}
\bibfield{author}{\bibinfo{person}{Anne-Kathrin Peters},
  \bibinfo{person}{Rafael Capilla}, \bibinfo{person}{Vlad~Constantin Coroamă},
  \bibinfo{person}{Rogardt Heldal}, \bibinfo{person}{Patricia Lago},
  \bibinfo{person}{Ola Leifler}, \bibinfo{person}{Ana Moreira},
  \bibinfo{person}{João~Paulo Fernandes}, \bibinfo{person}{Birgit
  Penzenstadler}, \bibinfo{person}{Jari Porras}, {and} \bibinfo{person}{Colin~C
  Venters}.} \bibinfo{year}{2024}\natexlab{}.
\newblock \showarticletitle{{Sustainability in Computing Education: A
  Systematic Literature Review}}.
\newblock \bibinfo{journal}{\emph{{ACM} Trans. Comput. Educ.}}
  \bibinfo{volume}{24} (\bibinfo{year}{2024}), \bibinfo{pages}{1--53}.
\newblock
Issue 1.
\urldef\tempurl%
\url{https://doi.org/10.1145/3639060}
\showURL{%
\tempurl}


\bibitem[Qiu et~al\mbox{.}(2024)]%
        {qiu2024power}
\bibfield{author}{\bibinfo{person}{Haoran Qiu}, \bibinfo{person}{Weichao Mao},
  \bibinfo{person}{Archit Patke}, \bibinfo{person}{Shengkun Cui},
  \bibinfo{person}{Saurabh Jha}, \bibinfo{person}{Chen Wang},
  \bibinfo{person}{Hubertus Franke}, \bibinfo{person}{Zbigniew~T. Kalbarczyk},
  \bibinfo{person}{Tamer Ba\c{s}ar}, {and} \bibinfo{person}{Ravishankar~K.
  Iyer}.} \bibinfo{year}{2024}\natexlab{}.
\newblock \showarticletitle{Power-aware deep learning model serving with
  µ-serve}. In \bibinfo{booktitle}{\emph{Proceedings of the 2024 USENIX
  Conference on Usenix Annual Technical Conference}} (Santa Clara, CA, USA)
  \emph{(\bibinfo{series}{USENIX ATC'24})}. \bibinfo{publisher}{USENIX
  Association}, \bibinfo{address}{USA}, Article \bibinfo{articleno}{5},
  \bibinfo{numpages}{19}~pages.
\newblock
\showISBNx{978-1-939133-41-0}


\bibitem[Rajput et~al\mbox{.}(2024a)]%
        {RajputMSR24}
\bibfield{author}{\bibinfo{person}{Saurabhsingh Rajput}, \bibinfo{person}{Maria
  Kechagia}, \bibinfo{person}{Federica Sarro}, {and} \bibinfo{person}{Tushar
  Sharma}.} \bibinfo{year}{2024}\natexlab{a}.
\newblock \showarticletitle{Greenlight: Highlighting TensorFlow APIs Energy
  Footprint}. In \bibinfo{booktitle}{\emph{Proceedings of the 21st
  International Conference on Mining Software Repositories}} (Lisbon, Portugal)
  \emph{(\bibinfo{series}{MSR '24})}. \bibinfo{publisher}{{ACM}},
  \bibinfo{address}{New York, NY, USA}, \bibinfo{pages}{304–308}.
\newblock
\showISBNx{9798400705878}
\href{https://doi.org/10.1145/3643991.3644894}{doi:\nolinkurl{10.1145/3643991.3644894}}


\bibitem[Rajput and Sharma(2024)]%
        {rajput2024benchmarking}
\bibfield{author}{\bibinfo{person}{Saurabhsingh Rajput} {and}
  \bibinfo{person}{Tushar Sharma}.} \bibinfo{year}{2024}\natexlab{}.
\newblock \showarticletitle{{ Benchmarking Emerging Deep Learning Quantization
  Methods for Energy Efficiency }}. In \bibinfo{booktitle}{\emph{2024 IEEE 21st
  International Conference on Software Architecture Companion (ICSA-C)}}.
  \bibinfo{publisher}{{IEEE}}, \bibinfo{address}{Los Alamitos, CA, USA},
  \bibinfo{pages}{238--242}.
\newblock
\href{https://doi.org/10.1109/ICSA-C63560.2024.00049}{doi:\nolinkurl{10.1109/ICSA-C63560.2024.00049}}


\bibitem[Rajput et~al\mbox{.}(2024b)]%
        {RajputTOSEM24}
\bibfield{author}{\bibinfo{person}{Saurabhsingh Rajput}, \bibinfo{person}{Tim
  Widmayer}, \bibinfo{person}{Ziyuan Shang}, \bibinfo{person}{Maria Kechagia},
  \bibinfo{person}{Federica Sarro}, {and} \bibinfo{person}{Tushar Sharma}.}
  \bibinfo{year}{2024}\natexlab{b}.
\newblock \showarticletitle{Enhancing Energy-Awareness in Deep Learning through
  Fine-Grained Energy Measurement}.
\newblock \bibinfo{journal}{\emph{ACM Trans. Softw. Eng. Methodol.}}
  \bibinfo{volume}{33}, \bibinfo{number}{8}, Article \bibinfo{articleno}{211}
  (\bibinfo{date}{Dec.} \bibinfo{year}{2024}), \bibinfo{numpages}{34}~pages.
\newblock
\showISSN{1049-331X}
\href{https://doi.org/10.1145/3680470}{doi:\nolinkurl{10.1145/3680470}}


\bibitem[Reddi et~al\mbox{.}(2020)]%
        {reddi2020mlperf}
\bibfield{author}{\bibinfo{person}{Vijay~Janapa Reddi},
  \bibinfo{person}{Christine Cheng}, \bibinfo{person}{David Kanter},
  \bibinfo{person}{Peter Mattson}, \bibinfo{person}{Guenther Schmuelling},
  \bibinfo{person}{Carole-Jean Wu}, \bibinfo{person}{Brian Anderson},
  \bibinfo{person}{Maximilien Breughe}, \bibinfo{person}{Mark Charlebois},
  \bibinfo{person}{William Chou}, \bibinfo{person}{Ramesh Chukka},
  \bibinfo{person}{Cody Coleman}, \bibinfo{person}{Sam Davis},
  \bibinfo{person}{Pan Deng}, \bibinfo{person}{Greg Diamos},
  \bibinfo{person}{Jared Duke}, \bibinfo{person}{Dave Fick},
  \bibinfo{person}{J.~Scott Gardner}, \bibinfo{person}{Itay Hubara},
  \bibinfo{person}{Sachin Idgunji}, \bibinfo{person}{Thomas~B. Jablin},
  \bibinfo{person}{Jeff Jiao}, \bibinfo{person}{Tom~St. John},
  \bibinfo{person}{Pankaj Kanwar}, \bibinfo{person}{David Lee},
  \bibinfo{person}{Jeffery Liao}, \bibinfo{person}{Anton Lokhmotov},
  \bibinfo{person}{Francisco Massa}, \bibinfo{person}{Peng Meng},
  \bibinfo{person}{Paulius Micikevicius}, \bibinfo{person}{Colin Osborne},
  \bibinfo{person}{Gennady Pekhimenko}, \bibinfo{person}{Arun Tejusve~Raghunath
  Rajan}, \bibinfo{person}{Dilip Sequeira}, \bibinfo{person}{Ashish Sirasao},
  \bibinfo{person}{Fei Sun}, \bibinfo{person}{Hanlin Tang},
  \bibinfo{person}{Michael Thomson}, \bibinfo{person}{Frank Wei},
  \bibinfo{person}{Ephrem Wu}, \bibinfo{person}{Lingjie Xu},
  \bibinfo{person}{Koichi Yamada}, \bibinfo{person}{Bing Yu},
  \bibinfo{person}{George Yuan}, \bibinfo{person}{Aaron Zhong},
  \bibinfo{person}{Peizhao Zhang}, {and} \bibinfo{person}{Yuchen Zhou}.}
  \bibinfo{year}{2020}\natexlab{}.
\newblock \showarticletitle{{ MLPerf Inference Benchmark }}. In
  \bibinfo{booktitle}{\emph{2020 ACM/IEEE 47th Annual International Symposium
  on Computer Architecture (ISCA)}}. \bibinfo{publisher}{{IEEE}},
  \bibinfo{address}{Los Alamitos, CA, USA}, \bibinfo{pages}{446--459}.
\newblock
\href{https://doi.org/10.1109/ISCA45697.2020.00045}{doi:\nolinkurl{10.1109/ISCA45697.2020.00045}}


\bibitem[Sarro(2023)]%
        {10260877}
\bibfield{author}{\bibinfo{person}{Federica Sarro}.}
  \bibinfo{year}{2023}\natexlab{}.
\newblock \showarticletitle{{ Search-Based Software Engineering in the Era of
  Modern Software Systems }}. In \bibinfo{booktitle}{\emph{2023 IEEE 31st
  International Requirements Engineering Conference (RE)}}.
  \bibinfo{publisher}{{IEEE}}, \bibinfo{address}{Los Alamitos, CA, USA},
  \bibinfo{pages}{3--5}.
\newblock
\href{https://doi.org/10.1109/RE57278.2023.00010}{doi:\nolinkurl{10.1109/RE57278.2023.00010}}


\bibitem[Schwartz et~al\mbox{.}(2020)]%
        {schwartz2020green}
\bibfield{author}{\bibinfo{person}{Roy Schwartz}, \bibinfo{person}{Jesse
  Dodge}, \bibinfo{person}{Noah~A. Smith}, {and} \bibinfo{person}{Oren
  Etzioni}.} \bibinfo{year}{2020}\natexlab{}.
\newblock \showarticletitle{Green AI}.
\newblock \bibinfo{journal}{\emph{Commun. ACM}} \bibinfo{volume}{63},
  \bibinfo{number}{12} (\bibinfo{date}{Nov.} \bibinfo{year}{2020}),
  \bibinfo{pages}{54–63}.
\newblock
\showISSN{0001-0782}
\href{https://doi.org/10.1145/3381831}{doi:\nolinkurl{10.1145/3381831}}


\bibitem[Selvan et~al\mbox{.}(2024)]%
        {selvan2025pepr}
\bibfield{author}{\bibinfo{person}{Raghavendra Selvan}, \bibinfo{person}{Bob
  Pepin}, \bibinfo{person}{Christian Igel}, \bibinfo{person}{Gabrielle Samuel},
  {and} \bibinfo{person}{Erik~B Dam}.} \bibinfo{year}{2024}\natexlab{}.
\newblock \showarticletitle{PePR: Performance Per Resource Unit as a Metric to
  Promote Small-Scale Deep Learning in Medical Image Analysis}.
\newblock  (\bibinfo{year}{2024}).
\newblock
\href{https://doi.org/10.48550/arXiv.2403.12562}{doi:\nolinkurl{10.48550/arXiv.2403.12562}}
\showeprint[arxiv]{2403.12562}
\newblock
\shownote{Presented at Northern Lights Deep Learning Conference 2025}.


\bibitem[Tedla et~al\mbox{.}(2024)]%
        {tedla_ecomls_2024}
\bibfield{author}{\bibinfo{person}{Meghana Tedla}, \bibinfo{person}{Shubham
  Kulkarni}, {and} \bibinfo{person}{Karthik Vaidhyanathan}.}
  \bibinfo{year}{2024}\natexlab{}.
\newblock \showarticletitle{{EcoMLS}: {A} {Self}-{Adaptation} {Approach} for
  {Architecting} {Green} {ML}-{Enabled} {Systems}}. In
  \bibinfo{booktitle}{\emph{2024 {IEEE} 21st {International} {Conference} on
  {Software} {Architecture} {Companion} ({ICSA}-{C})}}.
  \bibinfo{publisher}{IEEE}, \bibinfo{address}{Hyderabad, India},
  \bibinfo{pages}{230--237}.
\newblock
\showISBNx{9798350366259}
\href{https://doi.org/10.1109/ICSA-C63560.2024.00048}{doi:\nolinkurl{10.1109/ICSA-C63560.2024.00048}}


\bibitem[Verdecchia et~al\mbox{.}(2022)]%
        {verdecchia_data-centric_2022}
\bibfield{author}{\bibinfo{person}{Roberto Verdecchia}, \bibinfo{person}{Luis
  Cruz}, \bibinfo{person}{June Sallou}, \bibinfo{person}{Michelle Lin},
  \bibinfo{person}{James Wickenden}, {and} \bibinfo{person}{Estelle
  Hotellier}.} \bibinfo{year}{2022}\natexlab{}.
\newblock \showarticletitle{Data-{Centric} {Green} {AI} {An} {Exploratory}
  {Empirical} {Study}}. In \bibinfo{booktitle}{\emph{2022 {International}
  {Conference} on {ICT} for {Sustainability} ({ICT4S})}}.
  \bibinfo{publisher}{IEEE}, \bibinfo{address}{Plovdiv, Bulgaria},
  \bibinfo{pages}{35--45}.
\newblock
\showISBNx{978-1-66548-286-8}
\href{https://doi.org/10.1109/ICT4S55073.2022.00015}{doi:\nolinkurl{10.1109/ICT4S55073.2022.00015}}


\bibitem[Wohlin et~al\mbox{.}(2024)]%
        {Wohlin2024}
\bibfield{author}{\bibinfo{person}{Claes Wohlin}, \bibinfo{person}{Per
  Runeson}, \bibinfo{person}{Martin H\"{o}st}, \bibinfo{person}{Magnus~C.
  Ohlsson}, \bibinfo{person}{Bj\"{o}rn Regnell}, {and} \bibinfo{person}{Anders
  Wesslén}.} \bibinfo{year}{2024}\natexlab{}.
\newblock \bibinfo{booktitle}{\emph{Experimentation in Software Engineering}}.
\newblock \bibinfo{publisher}{Springer Berlin Heidelberg},
  \bibinfo{address}{Berlin}.
\newblock
\showISBNx{9783662693063}
\href{https://doi.org/10.1007/978-3-662-69306-3}{doi:\nolinkurl{10.1007/978-3-662-69306-3}}


\bibitem[{Wright} et~al\mbox{.}(2024)]%
        {wright2023efficiency}
\bibfield{author}{\bibinfo{person}{Dustin {Wright}}, \bibinfo{person}{Christian
  Igel}, \bibinfo{person}{Gabrielle Samuel}, {and} \bibinfo{person}{Raghavendra
  {Selvan}}.} \bibinfo{year}{2024}\natexlab{}.
\newblock \showarticletitle{Efficiency is Not Enough: A Critical Perspective of
  Environmentally Sustainable {AI}}.
\newblock  (\bibinfo{year}{2024}).
\newblock
\href{https://doi.org/10.1145/3724500}{doi:\nolinkurl{10.1145/3724500}}
\showeprint[arxiv]{2403.12562}~[cs.LG]


\bibitem[Xu et~al\mbox{.}(2025)]%
        {xu_edgellm_2025}
\bibfield{author}{\bibinfo{person}{Daliang Xu}, \bibinfo{person}{Wangsong Yin},
  \bibinfo{person}{Hao Zhang}, \bibinfo{person}{Xin Jin}, \bibinfo{person}{Ying
  Zhang}, \bibinfo{person}{Shiyun Wei}, \bibinfo{person}{Mengwei Xu}, {and}
  \bibinfo{person}{Xuanzhe Liu}.} \bibinfo{year}{2025}\natexlab{}.
\newblock \showarticletitle{{EdgeLLM}: {Fast} {On}-{Device} {LLM} {Inference}
  {With} {Speculative} {Decoding}}.
\newblock \bibinfo{journal}{\emph{IEEE Transactions on Mobile Computing}}
  \bibinfo{volume}{24}, \bibinfo{number}{4} (\bibinfo{date}{April}
  \bibinfo{year}{2025}), \bibinfo{pages}{3256--3273}.
\newblock
\showISSN{1536-1233, 1558-0660, 2161-9875}
\href{https://doi.org/10.1109/TMC.2024.3513457}{doi:\nolinkurl{10.1109/TMC.2024.3513457}}


\bibitem[Yarally et~al\mbox{.}(2023)]%
        {yarally2023batching}
\bibfield{author}{\bibinfo{person}{Tim Yarally}, \bibinfo{person}{Luis Cruz},
  \bibinfo{person}{Daniel Feitosa}, \bibinfo{person}{June Sallou}, {and}
  \bibinfo{person}{Arie van Deursen}.} \bibinfo{year}{2023}\natexlab{}.
\newblock \showarticletitle{{ Batching for Green AI - An Exploratory Study on
  Inference }}. In \bibinfo{booktitle}{\emph{2023 49th Euromicro Conference on
  Software Engineering and Advanced Applications (SEAA)}}.
  \bibinfo{publisher}{{IEEE}}, \bibinfo{address}{Los Alamitos, CA, USA},
  \bibinfo{pages}{112--119}.
\newblock
\href{https://doi.org/10.1109/SEAA60479.2023.00026}{doi:\nolinkurl{10.1109/SEAA60479.2023.00026}}


\bibitem[Zhao et~al\mbox{.}(2024)]%
        {zhao2024end}
\bibfield{author}{\bibinfo{person}{Yiyang Zhao}, \bibinfo{person}{Yunzhuo Liu},
  \bibinfo{person}{Bo Jiang}, {and} \bibinfo{person}{Tian Guo}.}
  \bibinfo{year}{2024}\natexlab{}.
\newblock \showarticletitle{CE-NAS: An End-to-End Carbon-Efficient Neural
  Architecture Search Framework}.
\newblock  (\bibinfo{year}{2024}).
\newblock
\showeprint[arxiv]{2406.01414}~[cs.LG]
\urldef\tempurl%
\url{https://arxiv.org/abs/2406.01414}
\showURL{%
\tempurl}


\end{thebibliography}

\end{document}